\documentclass[fleqn,usenatbib]{mnras}

\usepackage{newtxtext,newtxmath}

\usepackage[T1]{fontenc}
\usepackage{ae,aecompl}
\usepackage[utf8]{inputenc}

\usepackage{graphicx}	
\usepackage{bm,tensor,braket}
\usepackage{xcolor}
\usepackage{enumerate}
\usepackage{mathtools}
\usepackage[normalem]{ulem}

\title{No Slip gravity  in light of LISA standard sirens}
\author[A. Allahyari et al.] {
Alireza Allahyari, $^{1}$\thanks{E-mail:alireza.al@ipm.ir}
Rafael C. Nunes,$^{2}$\thanks{E-mail: rafadcnunes@gmail.com}
David F. Mota,$^{3}$\thanks{E-mail: d.f.mota@astro.uio.no}
\\
$^{1}$School of Astronomy, Institute for Research in Fundamental Sciences (IPM)
P. O. Box 19395-5531, Tehran, Iran\\
Department of Astronomy and High Energy Physics,
Faculty of Physics, Kharazmi University, Tehran, Iran\\
$^{2}$Divis\~ao de Astrof\'isica, Instituto Nacional de Pesquisas Espaciais, Avenida dos Astronautas 1758, S\~ao Jos\'e dos Campos, 12227-010, SP, Brazil\\
$^{3}$Institute of Theoretical Astrophysics, University of Oslo, P.O. Box 1029 Blindern, N-0315 Oslo, Norway
}
\date{Accepted XXX. Received YYY; in original form ZZZ}

\pubyear{2022}
\begin{document}
\label{firstpage}
\pagerange{\pageref{firstpage}--\pageref{lastpage}}
\maketitle

\begin{abstract}
Standard sirens (SS) are the gravitational wave analog of the astronomical standard candles, and can provide powerful information about the dynamics of the Universe up to very high $z$ values. In this work, we generate three mock SS catalogs based on the merger of massive black hole binaries  which are expected to be observed in the LISA operating frequency band. Then, we perform an analysis to test  modifications of general relativity (GR)  inspired by the No Slip gravity framework. We find that in the best scenarios, we can constrain the free parameters which quantify deviations from GR to 21\% accuracy, while the Hubble parameter can be simultaneously fit to 6\% accuracy. In combination with CMB information, we find a 15\% accuracy on the modified gravity free parameters and 0.7\% accuracy on the Hubble parameter. The SS events at very large cosmological distances to be observed in LISA band will provide a unique way to test  nature of gravity, but in the context of the analysis performed here, it will not be possible to distinguish the No Slip gravity from GR.
\end{abstract}



\section{Introduction}

The gravitational wave (GW) astronomy will provide unprecedented opportunity to test fundamental physics. Currently, more than 50 coalescing compact binaries events have already been observed during the three running stages of the LIGO/VIRGO \citet{LIGOScientific:2018mvr}, \citet{LIGO_01, LIGO_02} and  other dozens more should be observed in the coming years. Crucially, the observation of the GWs from neutron star binaries GW170817 and its associated electromagnetic counterpart GRB 170817A have marked  the dawn of multi-messenger cosmology \citet{LIGOScientific:2017vwq, Goldstein:2017mmi, Savchenko:2017ffs, LIGOScientific:2017zic,LIGOScientific:2017ync, DES:2016apb,Arcavi:2017xiz, Tanvir:2017pws}. Such events are dubbed as standard sirens (SS), the gravitational analogue of standard candles \citet{Holz:2005df, Palmese:2019ehe}. The wealth of science that joint detections  bring can not be probed with either messenger alone. Other counterparts like neutrino emissions and polarization features associated to this GW event can shed more light on the nature of the merger \citet{ANTARES:2017bia,Shakeri:2018qal}. The importance of such a detection is because the redshift and the location of the source are obtained with more precision breaking the degeneracies in the parameter space. In particular, one appealing application for the SS observations, is the possibility of utilizing events like these to estimate cosmological parameters \citet{Schutz:1986gp}. This provides one complementary probe to constrain cosmological parameters, since  the cosmological parameters are encoded in the luminosity distance provided from these events.

An independent measurement of the Hubble constant, $H_0$, using the standard siren approach results in $H_0=70^{12}_{-8}$ km s${}^{-1}$ Mpc${}^{-1}$ \citet{LIGOScientific:2017adf}. The errors for one event are still large. However, with the advent of third generation detectors like Einstein Telescope \citet{Maggiore:2019uih}, Cosmic Explorer \citet{Sathyaprakash:2019nnu} which are based on the earth and space-borne detectors like LISA \citet{LISA:2017pwj, Baker:2019nia}  and TianQin \citet{TianQin:2020hid,TianQin:2015yph}, the constraining power of GWs with electromagnetic counterparts  will increase significantly and they could  extend these observations to large redshifts. The LISA will operate in the millihertz band with the objective to be an all-sky GW survey \citet{Baker:2019nia}. Science with LISA brings opportunities and challenges in terms of complications which arise because of its motion around the earth. Basically, LISA can be considered as two detectors. It will be launched in three identical drag-free spacecraft forming an equilateral triangle with the arm length of about $2.5\times10^6 km$ \citet{Cutler:1997ta,Cornish:2002rt}.

The implications of cosmological studies using the standard sirens have motivated focused studies on the nature of dark energy, modified gravity, dark matter, and several other fundamental questions in modern cosmology \citet{Nishizawa:2019rra,LISACosmologyWorkingGroup:2019mwx,Belgacem:2018lbp,Belgacem:2017ihm, Ezquiaga:2021ler,Cai:2021ooo,Nishizawa:2017nef,Nunes:2019bjq,Yang:2019vni,Matos:2021qne,Jiang:2021mpd,Wang:2019tto,DAgostino:2019hvh,Pan:2021tpk,Kalomenopoulos:2020klp,Yang:2021qge,Gray:2019ksv,Lagos:2020mzy,Tasinato:2021wol,Bonilla:2019mbm,deSouza:2019ype,Dalang:2019rke,Zhang:2020axa,Yang:2020wby,Baker:2020apq,Fu:2020btp,Belgacem:2019zzu,Bonilla:2021dql,Mastrogiovanni:2020gua, Nunes:2020rmr,Baral:2020mzs,Bernardo:2021vsj,Mastrogiovanni:2020mvm,Wang:2021srv,Zhang:2021kqn, Canas-Herrera:2021qxs,Canas-Herrera:2019npr}. 
Even if the detected event has no electromagnetic counterpart, it is possible to use other methods to study cosmological parameters \citet{Mukherjee:2020mha,Zhu:2021aat,Garoffolo:2020vtd, Mukherjee:2020hyn,Borhanian:2020vyr,Wang:2020dkc,Feeney:2020kxk,Mastrogiovanni:2021wsd} 

In this article, we will forecast bounds for a motivated modified gravity model named No Slip gravity \citet{Linder:2018jil}, a subclass of the Horndeski gravity model \citet{Horndeski}, a general scalar tensor theory with second order field equations. The main characteristics of No Slip model are given by the speed of GWs propagation which is equal to the speed of light, and equality between the effective gravitational coupling strengths to matter and light, but yet different from Newton’s constant, which is capable of generating an effective gravitational coupling. Some observational perspectives of this class were previously investigated in \citet{Brush:2018dhg, Mitra:2020vzq,Linder:2020xza,Brando:2019xbv,Nunes:2020rmr}.  
Our approach is to use the LISA standard sirens to predict the bounds on the No Slip gravity baseline parameters.  In following analysis, we integrate the modified luminosity distance from standard sirens to constrain the main free parameters of the model. Also, we combine our results with previous results where cosmic microwave background data were used for No Slip gravity, to get tight constraints on the model.

This manuscript is organized as follows. In section \ref{2}, we present the modified luminosity distance in our model. Section \ref{sec:data} is devoted to the LISA standard sirens where we show how we generate the mock data. In section \ref{results}, we present our main results. Finally, our final remarks are included in section \ref{5}.

\section{Propagation of Gravitational Waves in Modified Gravity}
\label{2}
One particular way by which modified theories leave their imprints on the cosmological observables is  the modified propagation equations for the tensorial part of the perturbations. This leads to a notion of GW luminosity distance different from the electromagnetic luminosity distance. This provides an arena to test the modified theories in the context of GWs by the standard sirens, the GW events with the associated electromagnetic counterparts.
The standard expression of the luminosity distance in a universe with matter density fraction $\Omega_m$, radiation density fraction $\Omega_R$ and dark energy density $\rho_{de}$  is defined as 
\begin{equation}
 d^{em}_L(z)=\frac{1+z}{H_0}\int_0^z\, 
\frac{d\tilde{z}}{E(\tilde{z})}\, ,
\end{equation}
where
\begin{equation}
\label{E(z)}
E(z)=\sqrt{\Omega_R (1+z)^4+\Omega_m (1+z)^3+\rho_{de}(z)/\rho_0 }\, ,
\end{equation}
where we have  $\rho_0=3H_0^2/(8\pi G)$ \citet{Belgacem:2017ihm}.
In GR the transverse taceless part of metric perturbations, $\partial_{i}h^{ij}=h^{i}_{i}=0$, produces the gravitational waves. The propagation equation for GWs on a flat FRW background in absence of anisotropic stress tensor  for the polarization A  is given by
\begin{align}
h''_A  +2 {\cal H} h'_A+k^2 h_A=0, \quad\, A=+,\times 
\end{align}
where the prime denotes derivative with respect to the conformal time and ${\cal H}=a'/a$.
Defining
\begin{align}
h_{A}(\eta,k)=\frac{\chi_{A} (\eta,k)}{a(\eta)}\, ,
\end{align}

we find that
\begin{align}
\chi_{A}''(\eta,k)+\left(k^2-\frac{a''}{a} \right)\chi_{A}(\eta,k)=0 \,.
\end{align}
This shows that the amplitude of gravitational waves decrease as $1/a$ and in GR the luminosity distance for the gravitational waves from a binary is the same as electromagnetic waves $d_{L}^{em}=d_{L}^{gw}$.
In a generic modified theory the equation for GWs in the Fourier space  reads as
\begin{align}
h''_A  +2 {\cal H}[1-\delta(\eta)] h'_A+k^2 h_A=0\, .
\label{newdis}
\end{align} 

We have neglected the modulations in  $k^2$ term as they change the speed of GWs and we assume the vacuum field equations. Changing the speed of GWs is not favored given the stringent limit set by GW170817  $|c_{\rm gw}-c|/c< O(10^{-15})$.
To solve Eq.~\eqref{newdis} one introduces
\begin{align}
h_{A}(\eta,k)=&\frac{\chi_{A} (\eta,k)}{\bar{a}(\eta)}\,,\\ 
\frac{\bar{a}'}{\bar{a}}=&{\cal H} \left(1-\delta(\eta) \right). 
\end{align}

This way one finds that the luminosity distance for the GWs obeys a different equation. We have \citet{Belgacem:2017ihm, Finke:2021aom, Mastrogiovanni:2020mvm,Hogg:2020ktc}
\begin{align}
d_L^{\,\rm gw}(z)=d_L^{\,\rm em}(z)\exp\left\{-\int_0^z \,\frac{dz'}{1+z'}\,\delta(z')\right\}\, .
\label{dis}
\end{align}

The modified gravity we consider which changes the propagation of GWs as we have in Eqs.~\eqref{newdis} and ~\eqref{dis} is the No Slip model. This modified theory has the advantage of not changing the propagation speed for GWs. Therefore, it is favoured by the current observations. This model exhibits simple behavior in terms of phenomenological parametrizations and has been applied to study the CMB \citet{Brush:2018dhg}.

No Slip model belongs to the well-motivated larger class of Horndeski scalar-tensor theories \citet{Horndeski}. The Horndeski is the most general scalar-tensor theory with second order equations of motion in four dimensions from which different modified theories can be obtained such as Quintessence, k-essence and $f(R)$ gravity. The Lagrangian for the theory is presented in Refs. \citet{Kobayashi:2019hrl,Gleyzes:2014rba} and other references therein. This theory can be tuned such that GWs propagate at the speed of light to conform with  the limit set by GW170817. Moreover, a very important prediction of modified gravity theories is the modification of Einstein equations in a way where the coupling of gravity to matter and light could be different. In other words, the Poisson equation and  the lensing equation have different  dependence on the metric perturbations. 
The perturbations on the FRW background in the conformal Newtonian gauge are given by
\begin{align}
ds^2=a(\eta)^2\left\lbrace -\left(1+2 \Psi \right)d\eta^2+\left(1-2\Phi \right)\delta_{ij}dx^{i}dx^j   \right\rbrace ,
\end{align}
where $\Phi$ and $\Psi$ are metric perturbations. It is  usually a combination, $\Phi+\Psi$, which appears in the geodesic equations for photons \citet{Bertacca:2014wga,Allahyari:2017gmq}. More explicitly, the Einstein equations take the form
\begin{align}
 \nabla^{2} & \Psi = 4\pi a^{2} G_N G_{\rm matter} \rho_m\,\delta_ m \\  
\nabla^{2} & (\Psi+\Phi) = 8\pi a^{2} G_N G_{\rm light} 
\rho_m\,\delta_m \,,
\end{align}
where the quantity $\rho_m$ is the background matter density, and 
$\delta_m=\delta\rho_m/\rho_m$ is the comoving density perturbation and $G_{\rm light}$ and $ G_{\rm matter}$ are functions of time and $G_N$ is the gravitational constant.

To quantify the effect of gravity on matter and photons the gravitational slip parameter is  defined by 
\begin{align}
\eta=\frac{G_{matter}}{G_{light}}.
\end{align}

In the No Slip model the gravitational interaction is gauged so that for the gravitational slip parameter we have $\eta=1$. This way, we find that  the effective gravitational couplings are simply related by 
\begin{align}
G_{matter}=G_{light}=\frac{m_{p}^2}{M_{*}^{2}}\,,
\end{align}
where $m_{p}$ and $M_{*}$ are Planck and time-dependent Planck mass respectively.  $M_{*}$ is a function of the terms in the Horndeski Lagrangian \citet{Kase:2018aps}, \citet{Linder:2015rcz}. We find that the gravitational coupling strength is modified. Therefore, the growth of structures and lensing potentials will deviate from GR prediction.

The other condition to impose on the Horndeski model to derive No Slip model is the speed of GWs. The Hornedeski model can accommodate different propagation speeds for GWs than the speed of light. In Horndeski model the second order action for tensorial perturbations reads as
\begin{align}
S^{(2)}_{\rm tensor}=
\frac{1}{8}\int  a^3 q_t\left[
\dot h_{ij}^2-\frac{c_{gw}^2}{a^2}(\partial_k h_{ij})^2
\right] dtd^3x,
\label{action2:tensor}
\end{align}
where $q_t$ is defined in Ref.~\citet{Kase:2018aps}. The speed of tensorial perturbations is given by $c_{gw}$ which is function of various terms in the Horndeski Lagrangian.

In No Slip model we impose the condition to have $c_{gw}=1$. Varying the action in Eq.~\eqref{action2:tensor} and imposing the conditions mentioned, we get
\begin{align}
    \ddot{h}_A+\left(3H +\frac{\dot{q}_t}{q_t} \right) \dot{h}_A
+ \frac{k^2}{a^2} h_A=0\,.
\label{tensor}
\end{align}
This way we find that $G_{matter}=G_{light}=1/(16\pi G_N G_4)$ and $ q_t=2 G_4=M_*^2$ where $G_4$ is one of the terms in the Horndeski Lagrangian~ \citet{Kase:2018aps}.
Comparing Eqs.~\eqref{tensor} and ~\eqref{newdis}, we have $-2\delta=\frac{d \ln{q_t}}{d \ln{a}}$. We find that it is the running of the Planck mass which controls the tensor perturbations in this theory.
At this point, one could define a simple phenomenological parametrization to capture the main features of the theory. One could start with a parametrization for $\delta$ or $M_*$. As we are interested in studying the GW luminosity distance we take the first appraoch, namely $\delta$ parametrization.
Specializing to our case study here, No Slip model, we also use the phenomenological parametrization introduced by \citet{LISACosmologyWorkingGroup:2019mwx}
\begin{align}
\label{delta2}
\delta(z) = \frac{n(1-\xi)}{1-\xi+\xi(1+z)^n}\;,
\end{align}
where $\xi$ and $n$ are constants. This parametrization is applicable to models  in which we expect some simple properties in low and high redshifts. More explicitly, we find that at the high redshifts  $z\rightarrow \infty$, we have $\delta(z)\rightarrow 0$. Moreover, when $z\rightarrow0$, this yields $\delta(z)\rightarrow n(1-\xi)$ where for $\xi=1$ we have $\delta(z)\rightarrow 0$. Thus, we recover the standard expression for the luminosity distance at high and low redshifts. We expect this parametrization to be a viable parametrization as theory is not complicated. Please also note that in models where there are also other degrees of freedom other than one scalar field, the parametrization in Eq.~\eqref{delta2} might not capture the details of the theory. Examples where this parametrization might not work are bigravity theories \citet{Comelli:2012db,Hassan:2011zd}.
Alternatively, another parametrization starting with variation of Planck mass is given in Refs. \citet{Brush:2018dhg, Mitra:2020vzq},
\begin{align}
\left( \frac{m_{p}}{M_{*}}\right)^{-2}=1+\frac{\mu}{1+(\frac{a}{a_{t}})^{-\tau} } \,,
\label{linder}
\end{align}
where $\mu$ is the amplitude of transition, $a_t$ is the scale factor at the transition time and $\tau$ is the rapidity. This parametrization has a simple transition from unity in the past to some constant value in the future. We find that simple parametrizations of the model are adequate for No Slip model.
Here, we find it more convenient to work in $n,\xi$ parametrization, as defined in  Eq.~(\ref{delta2}). For the sake of comparison, the mapping from set $(n,\xi)$ to set $(\tau,\mu)$ has been provided in Ref.~\citet{Mitra:2020vzq} 
\\
\begin{align}
 &\xi =  \lim_{z\to \infty}\frac{M_{*}(0)}{M_{*}(z)} \,, \label{eq:Xi0param} \\
& n  \simeq \frac{\alpha_{M0}}{2(\xi-1)} \,,
\label{eq:nparam}
\end{align}
where $\alpha_{M0}=-2\delta(0)$. One could check that the mapping  generates  the same behavior for $\delta(z)$ in two parametrizations. The parametrization in Eq.~\eqref{linder} was suggested to produce a stable and ghost free model. Therefore, this mapping ensures that parametrization in Eq.~\eqref{delta2} is a good choice.

\section{LISA Standard Sirens}
\label{sec:data}

LISA is a space-borne detector with its sensitivity peak around 1 millihertz. Among astrophysical sources LISA can  reach including Galactic binaries \citet{Breivik:2017jip,Korol:2017qcx}, stellar-origin black hole binaries \citet{Sesana:2016ljz} and  extreme-mass-ratio inspirals \citet{Babak:2017tow}, LISA will also observe massive black hole binaries (MBHBs) from $10^4$ to $10^7$ solar masses \citet{Klein:2015hvg}. The high signal to noise ratio (SNR) of the detected signals will allow for more precise parameter estimations. Among the most probable LISA sources with electromagnetic counterparts are MBHBs.  In particular, MBHBs are  supposed to merge in  gas rich environments and within the LISA frequency band allowing for electromagnetic followups to determine their redshifts. The prospect of  MBHBs which could have EM counterparts extends  up to $z$ $\sim$7 providing  a unique probe of the universe at high redshifts \citet{Klein:2015hvg}.

Our catalog is based on the model in Ref.~\citet{Tamanini:2016zlh} where they use a semi analytic model which allows tracing the galactic baryonic structures and dark matter mergers. In addition, the model integrates the black hole seeding at high redshifts and the delays between the merger
of two galaxies and that of the massive BHs residing in the galaxies. 

Theoretical models and simulations can predict the redshift distribution and merger rate of MBHBs. Depending on the initial conditions for blackhole formation at high redshifts, there are two scenarios light seed and heavy seed. In the light seed scenario massive BHs are assumed to grow from the remnants of
population III (pop III) stars forming at $z\approx15-20$. In the heavy seed scenario, the massive BHs are assumed to form from the collapse  of protogalactic disks \citet{Tamanini:2016zlh}.
The result of the  scenarios produce three categories of population models named Pop III, Delay and No Delay \citet{Madau:2001sc,Volonteri:2007ax,Klein:2015hvg}. Among all the merger events, one chooses the events with a SNR threshold and localization error $\Delta\Omega$.
The redshift distribution of sources based on LISA sensitivity is reported in Ref.~\citet{Tamanini:2016zlh,Tamanini:2016uin}.

For each source one then calculates  the associated  GW luminosity distance assuming a given cosmology. We need to consider the prospective errors on $d_L$ based on the LISA sensitivity. Using the criteria in Ref. \cite{Speri:2020hwc} where they use the results of Ref.~\citet{Marsat:2020rtl} to fix the proportionality factor in the approximation $\sigma_{\rm LISA} / d_L \propto 2 d_L$ , we have
\begin{equation}
\frac{\sigma_{\rm LISA}(z)}{d_L(z)} = 0.05 \left(\frac{d_L(z)}{36.6\, {\rm Gpc}}\right) \,.
\end{equation}

However, this will give us the instrument error. We also need to consider three pieces of ingredients in our error estimation \citet{Wang:2021srv, Speri:2020hwc}. They are weak lensing
\begin{equation}
\sigma_{\rm delens}(z) = F_{\rm delens}(z) \sigma_{\rm lens}(z) \,,
\end{equation}
where we have
\begin{equation}
F_{\rm delens}(z) = 1 - \frac{0.3}{\pi / 2} \arctan\left(\frac{z}{z_*}\right) \,,
\end{equation}
\begin{equation} 
\frac{\sigma_{\rm lens}(z)}{d_L(z)} = 0.066 \left( \frac{1 - (1+z)^{-0.25}}{0.25} \right)^{1.8} \,,
\end{equation}
peculiar velocity
\begin{equation}
\frac{\sigma_v(z)}{d_L(z)} = \left[ 1 + \frac{c (1+z)^2}{H(z)d_L(z)} \right] \frac{\sqrt{\langle v^2 \rangle}}{c} \,,
\end{equation} 
with  $\langle v^2 \rangle = 500$ km/s,
and redshift measurement error \citet{Speri:2020hwc}
\begin{equation}
\sigma_{\rm photo}(z) = 0.03 (1+z) \quad{\rm if}\; z>2 \,.
\end{equation} 

The total error is given by adding in quadrature all contributions:
\begin{equation}
\langle\sigma_{d_L}  \rangle^2=\langle \sigma_\mathrm{photo} \rangle^2+\langle \sigma_{lens} \rangle^2+\langle \sigma_{LISA} \rangle^2+\langle \sigma_{v} \rangle^2\;.
\end{equation}

We generate three catalogs based on these criteria for the aforementioned distributions, namely, Pop III, Delay and No Delay for a ten-year mission. We generate our mock catalog from a fiducial input baseline with $H_0=67.4$ km s${}^{-1}$ Mpc${}^{-1}$, $\Omega_m=0.315$, $\xi =  1.0328$, $n = 0.3176$. The pair values $(\xi,n)$ are chosen in order to generate a mock catalog using modified gravity corrections. The motivation for this choice meets in the sense that we do not  assume the generation of the full simulated catalogs fixed on the $\Lambda$CDM baseline. Since we are interested in investigating deviations from the standard cosmology, we choose some minimal input deviation from $\Lambda$CDM  cosmology to create the mock data. Then, the forecast of experiments, in this case, the LISA mission, will quantify the expected error bar on the baseline of the model. Note that our choice for the pair values $(\xi,n)$ is compatible with current constraints \citet{Brush:2018dhg}, \citet{Mitra:2020vzq}. 



\begin{figure}
	\begin{center}
		\includegraphics[width=1\columnwidth]{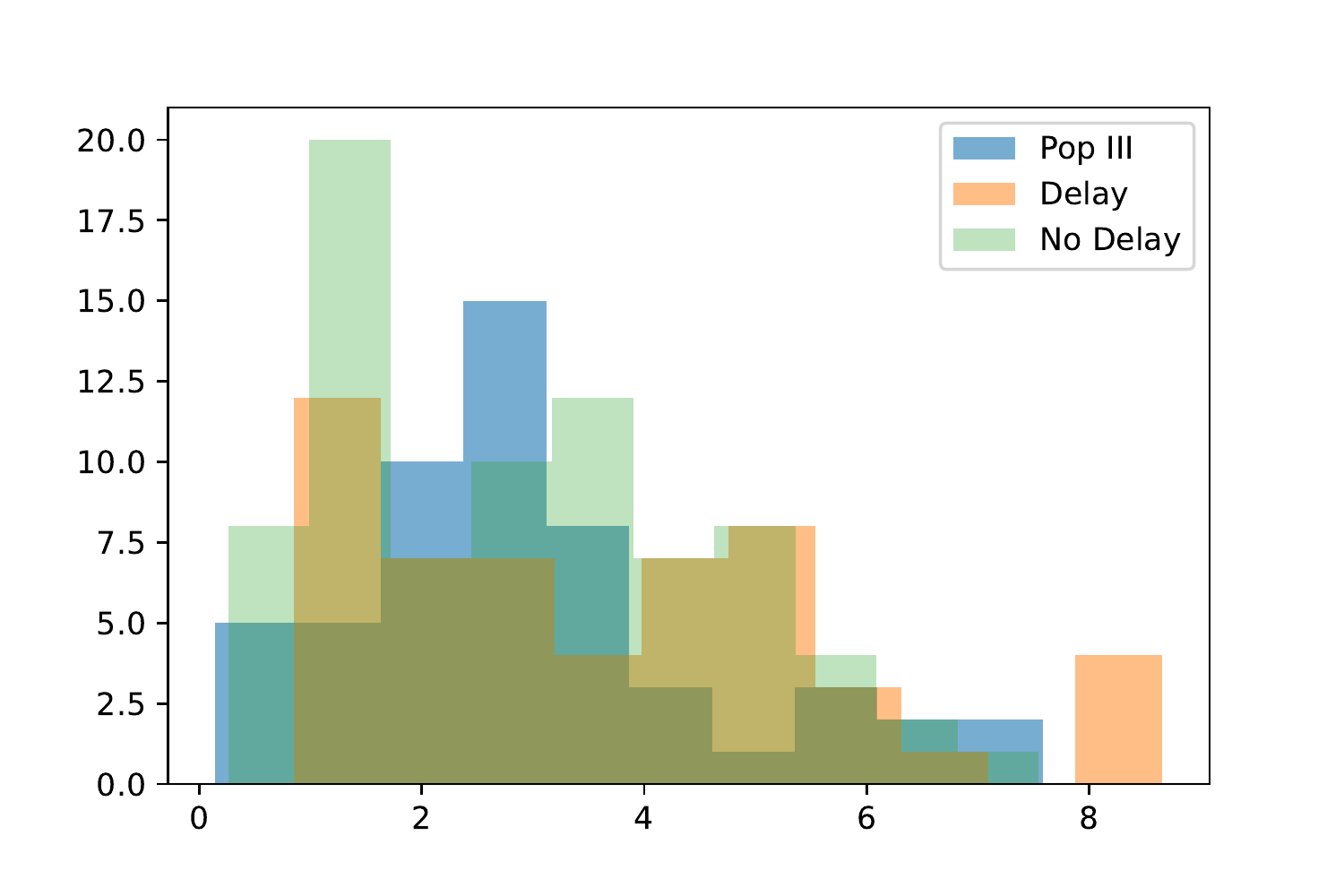}					
		\caption{Redshift distribution of MBHB standard sirens, namely, Pop III, Delay and No Delay.}	
		\label{z-dist}
	\end{center}
\end{figure}

\begin{figure}
	\begin{center}
		\includegraphics[width=1\columnwidth]{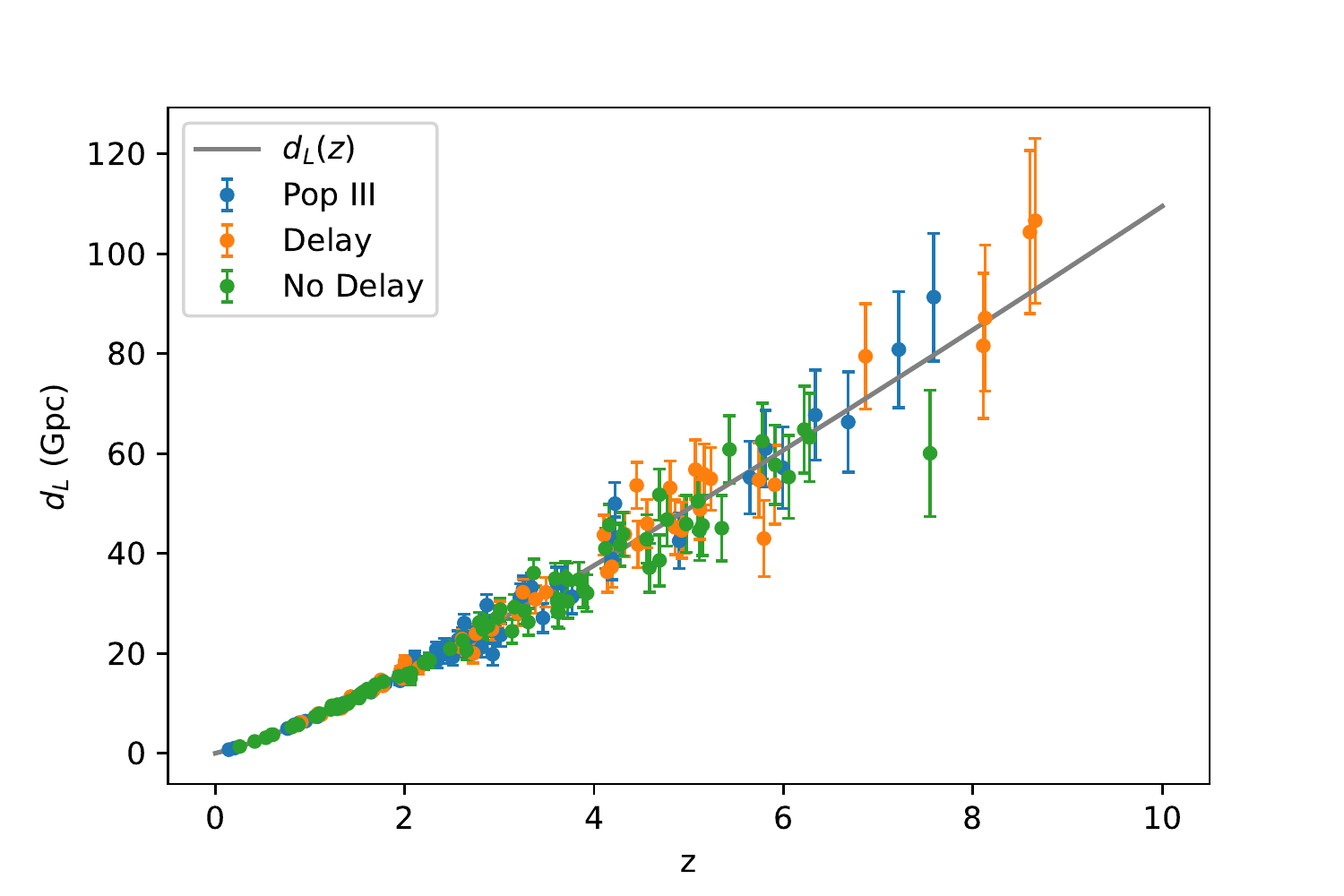}
		\caption{LISA standard sirens at all redshifts from the three population models we consider in this work.}	
		\label{SH1}
	\end{center}
\end{figure}

Figure \ref{z-dist} shows the redshift distributions of the mock catalogs and Figure \ref{SH1} shows the luminosity distance and the corresponding errors based on our input baseline, as well as the three categories of LISA SS sources. 
In what follows, we will analyze the free parameters of the theory and present our main results.

\section{Results and Discussion}
\label{results}

In this section, we present our main results on the model parameters by performing Bayesian Monte Carlo Markov Chain analysis. We create new likelihoods in \texttt{MontePython} code \citet{monte} to analyze the SS mock data as presented above. Then, we use the Metropolis - Hastings mode in \texttt{CLASS}+\texttt{MontePython} code~\citet{class, monte, monte_V3} to derive the constraints on cosmological parameters ensuring a Gelman - Rubin convergence criterion of $R - 1 < 10^{-3}$~\citet{gelman1992}. We divide our analysis as follows: First, we will analyze our baseline for each category of populations as defined in section \ref{sec:data}, namely, Pop III, Delay and No Delay samples. Then, we will do a joint analysis using the CMB information as obtained in the context of No Slip gravity in Ref.~\citet{Brush:2018dhg}. We simulate LISA + CMB constraints on the No Slip gravity adding appropriate Gaussian priors on the parameters $H_0$ and $w_{\rm cdm}$ as derivatives in Ref.~\citet{Brush:2018dhg}.

\begin{table*}
	\centering
	\caption{Constraints at 68\% CL from the Pop III samples and their combination with CMB information.}
	\label{tab:main_results_1}
	\begin{tabular}{cccc}
		\hline
		Parameter &  Pop III & Pop III + CMB \\
		\hline
		$H_0$ [km s${}^{-1}$ Mpc${}^{-1}$]  & $66.71\pm 0.95$ & $ 66.92\pm 0.56$\\
		
		$\Omega{}_{m}  $ & $0.31\pm 0.11   $ & $ 0.3160\pm 0.0070     $\\
		
		$n              $ & $0.64^{+0.56}_{-0.68}$ & $0.15^{+0.69}_{-0.80}   $\\
		
		$\xi            $ & $0.98\pm 0.33 $ & $ 1.016^{+0.069}_{-0.083} $\\
		
	\end{tabular}
\end{table*}

\begin{table*}
	\centering
	\caption{Same as table \ref{tab:main_results_1}, but for the Delay samples.}
	\label{tab:main_results_2}
	\begin{tabular}{cccc}
		\hline
		Parameter &  Delay & Delay + CMB \\
		\hline
		$H_0$ [km s${}^{-1}$ Mpc${}^{-1}$] & $70.8\pm 4.5 $ & $67.78\pm 0.91      $ \\
		
		$\Omega{}_{m } $ & $0.456^{+0.049}_{-0.13} $ & $0.3098\pm 0.0092$  \\
		
		$n  $ & $0.98\pm 0.55$ & $0.42^{+0.47}_{-0.78}$ \\
		
		$\xi$ & $1.362^{+0.081}_{-0.22}$ & $1.13^{+0.14}_{-0.34} $ \\
		
	\end{tabular}
\end{table*}

\begin{table*}
	\centering
	\caption{Same as table \ref{tab:main_results_1}, but for the No Delay samples.}
	\label{tab:main_results_3}
	\begin{tabular}{cccc}
		\hline
		Parameter &  No Delay & No Delay + CMB \\
		\hline
		$H_0$ [km s${}^{-1}$ Mpc${}^{-1}$]& $66.87^{+0.92}_{-0.63}$ & $67.21\pm 0.47      $ \\
		
		$\Omega{}_{m }  $ & $0.238\pm 0.078  $ & $0.3133\pm 0.0062      $ \\
		
		$n              $ & $0.43^{+0.37}_{-0.55} $ & $0.07^{+0.71}_{-0.54}   $ \\
		
		$\xi            $ & $0.75\pm 0.31$& $1.037^{+0.076}_{-0.17}$ \\
		
	\end{tabular}
\end{table*}

\begin{figure}
	\begin{center}
		\includegraphics[width=1\columnwidth]{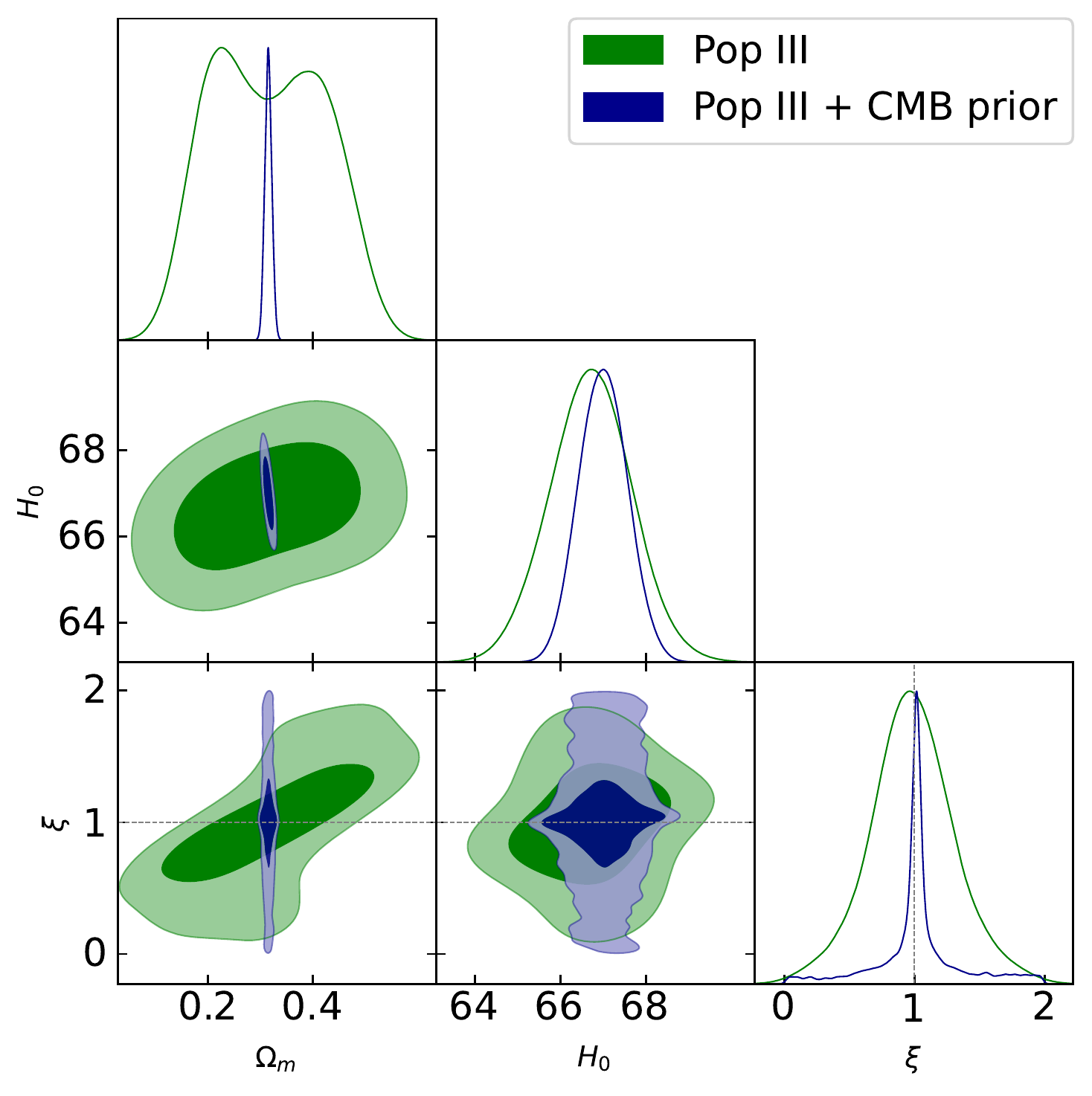}			
		\caption{One dimensional posterior distributions and two dimensional joint contours for the parameter space $H_0$, $\Omega_m$ and $\xi$ for LISA Pop III samples and Pop III + CMB prior. The $\xi =1$ prediction recovers the GR.}
		\label{LISA_I}
	\end{center}
\end{figure}

\begin{figure}
	\begin{center}
		\includegraphics[width=1\columnwidth]{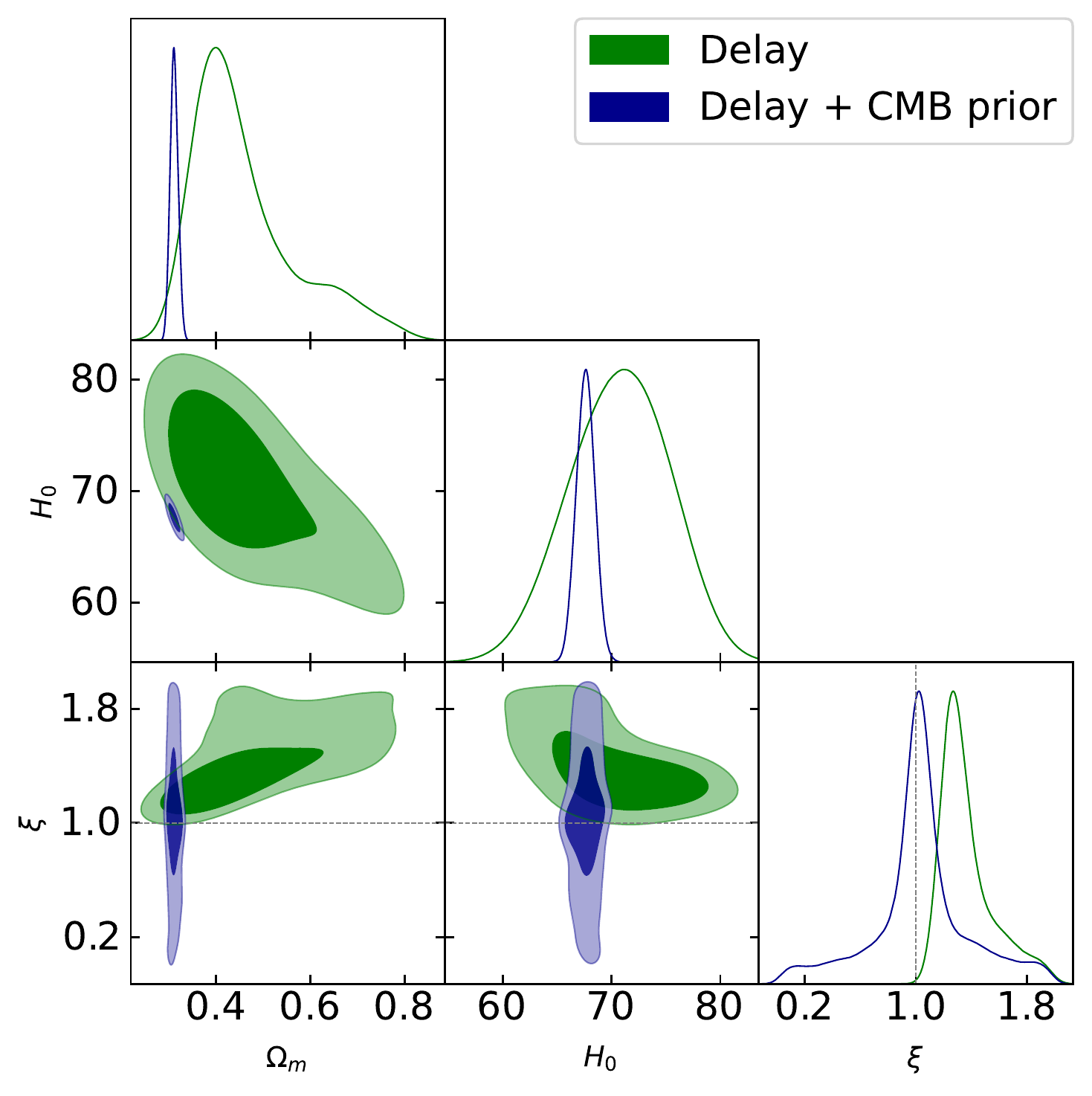}			
		\caption{Same as in Figure \ref{LISA_I}, but for the Delay samples.}
		\label{LISA_II}
	\end{center}
\end{figure}

\begin{figure}
	\begin{center}
		\includegraphics[width=1\columnwidth]{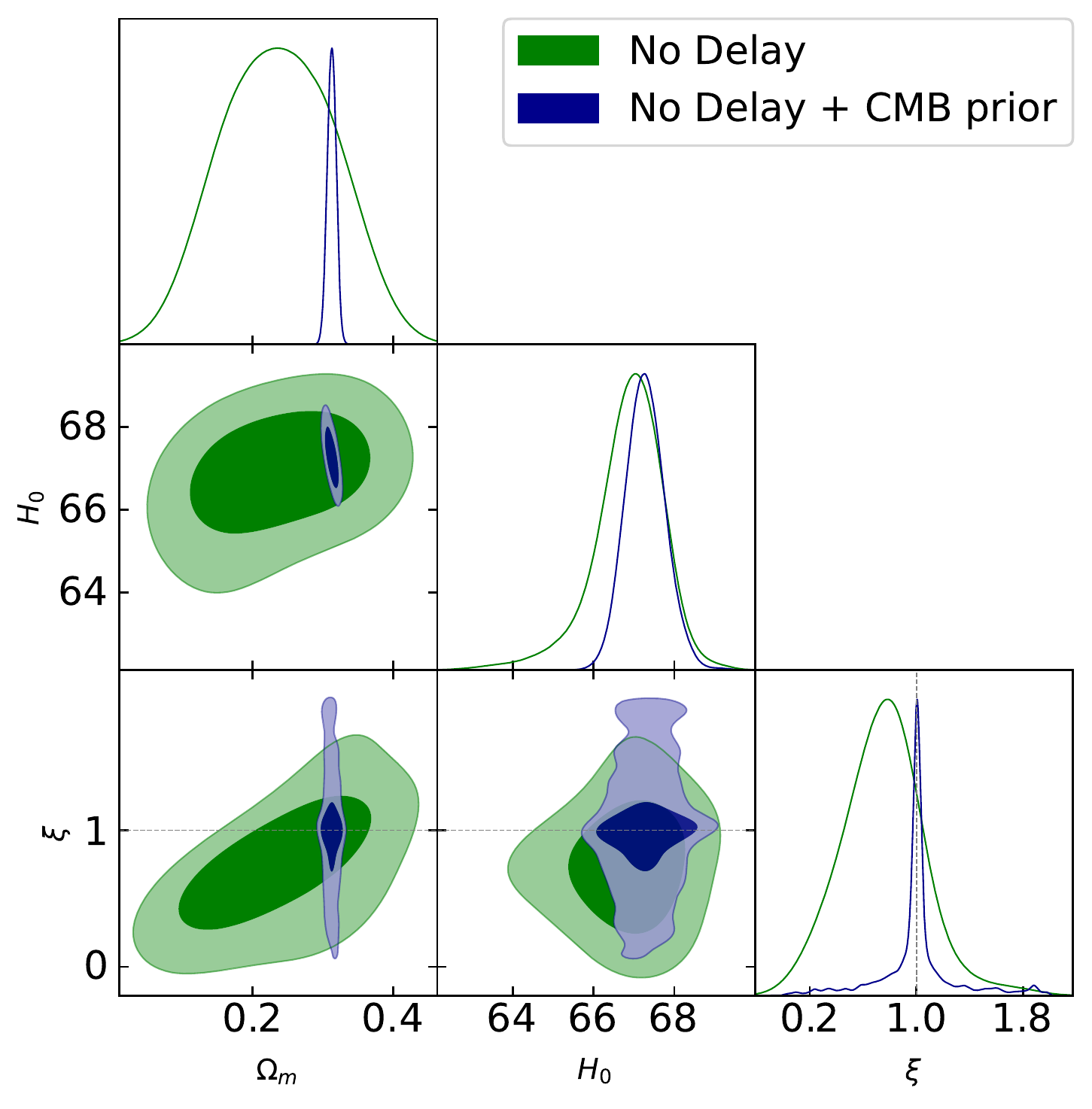}			
		\caption{Same as in Figure \ref{LISA_I}, but for the No Delay samples.}
		\label{LISA_III}
	\end{center}
\end{figure}

\begin{figure*}
\includegraphics[width=8cm]{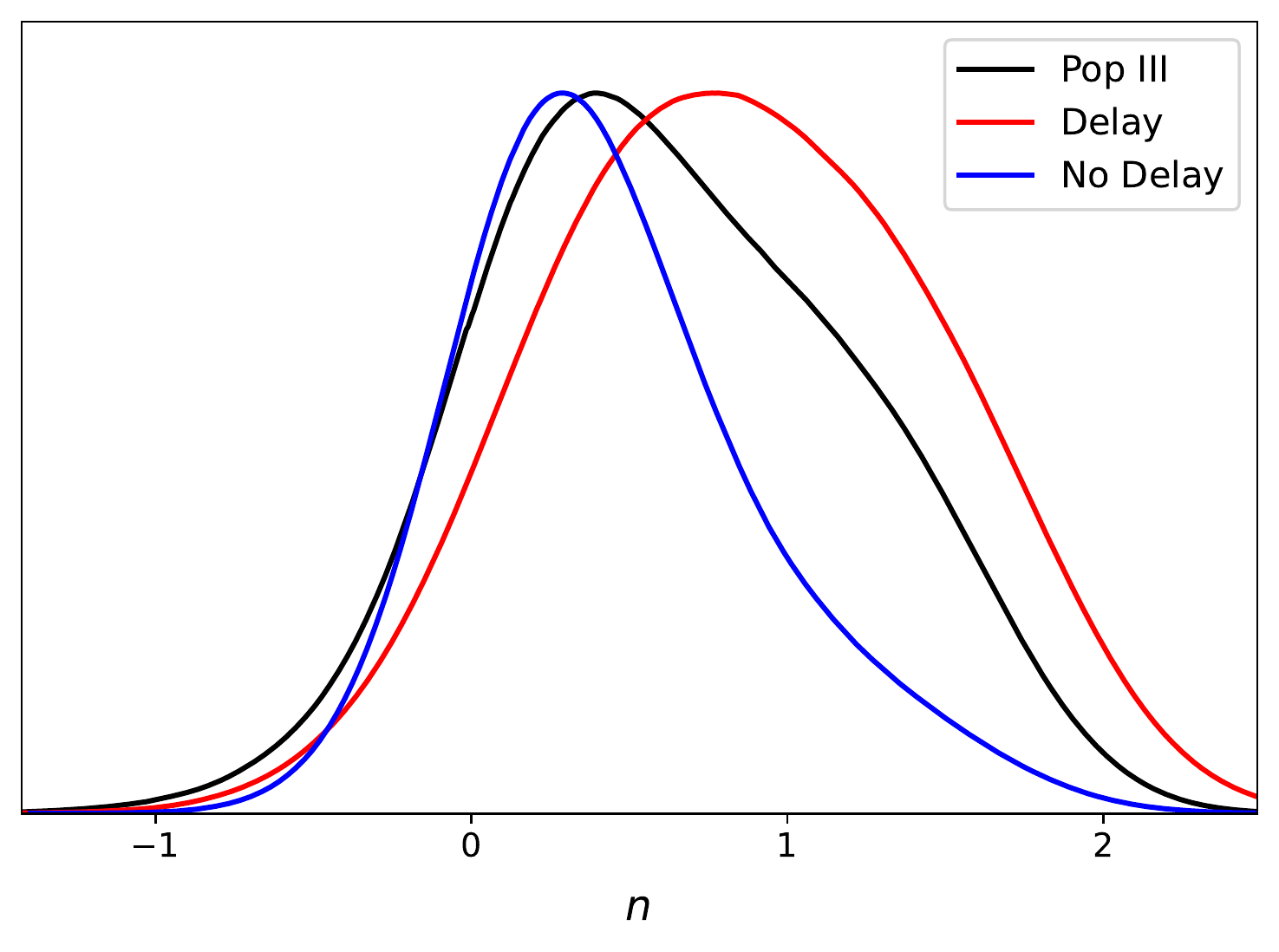} \,\,\,\,\,\,\,
\includegraphics[width=8cm]{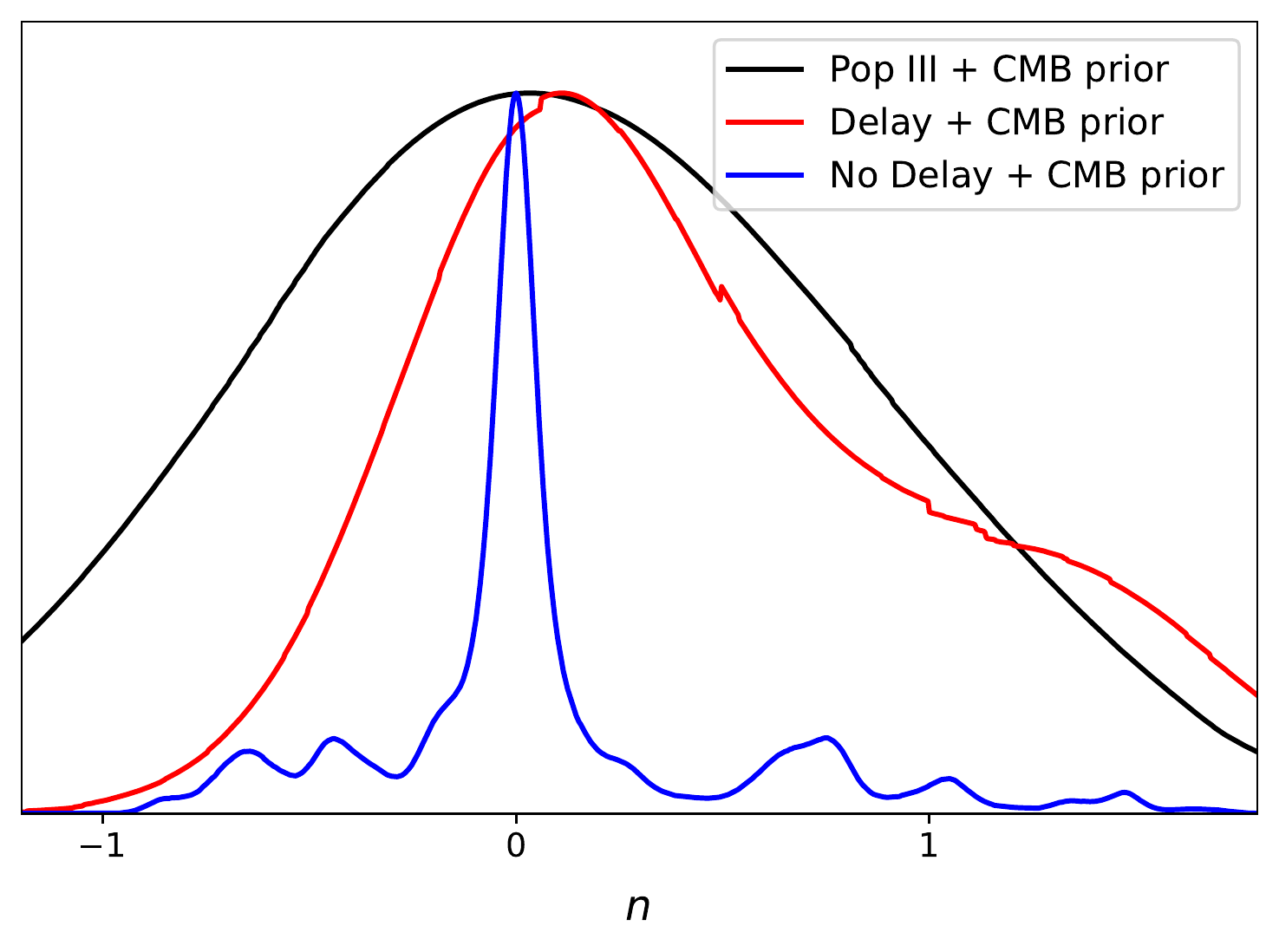}
\caption{1D marginalized posterior probability distributions for $n$, obtained for each LISA mock sample on the left panel and in combination with CMB prior on the right panel.}
\label{n}
\end{figure*}

We summarize the main results of our statistical analysis in tables \ref{tab:main_results_1}, \ref{tab:main_results_2} and \ref{tab:main_results_3} for the cases with Pop III, Delay and No Delay mock data, respectively.  The tables also show the best fit for LISA + CMB prior combination. Figures \ref{LISA_I}, \ref{LISA_II} and \ref{LISA_III} show the 2D joint posterior distributions at 68\%~CL and 95\%~CL for some parameters of interest. As expected by the definition in Eq. (\ref{delta2}), we note that the parameters $n$ and $\xi$ are completely degenerate statistically. In all our analysis, we choose to keep both parameters free, although only the parameter $\xi$  would suffice to quantify the deviations of GR. Figure \ref{n} shows the one dimensional marginalized posterior probability distributions for the parameter $n$.

We find that the free parameter which quantifies deviation of GR, $\xi$, can be measured with 33\%, 21\% and 41\% accuracy from the Pop III, Delay and No Delay populations, respectively. As already pointed out in \citet{Mitra:2020vzq}, we see a significant degeneracy between the parameters $\xi$ and $n$, which on the other hand is also very degenerate with the matter density $\Omega_m$ and the Hubble constant, $H_0$. Note that for $n=0$, or equivalently $\xi=1$, the  equations of the model also reduce to GR. It is interesting to note that the parameter $\xi$ is positively correlated with $\Omega_m$, and does not show significant correlation with $H_0$. As the matter density increases, the luminosity distance to the different GWs sources will decrease, but this can be balanced by increasing the effects from the modified gravity model. We find that the Hubble parameter can be simultaneously fit to 1.4\%, 6.2\% and 1.1\% accuracy from the Pop III, Delay and No Delay populations, respectively.

To test gravity through the propagation of gravitational waves is a unique probe, though the model cannot be distinguished due to strong degeneracy in the whole parameter space using only the SS events. Therefore, in this case, it will be important to combine the SS sample with other probes.  Geometrical probes measured by electromagnetic waves like Supernovae Type Ia, Baryon Acoustic Oscillation distance, cosmic chronometers, among others, are insensitive on the No Slip gravity free parameters, but can provide independent constraints on the $\Omega_m$ and $H_0$, and then to help break this degeneracy on $\xi$ and $n$. On the other hand, the No Slip gravity prediction is sensitive to the amplitude of late-time matter fluctuations (weak lensing and redshift-space distortions measurements) and CMB anisotropies. As defined previously, we combine the MBHB standard sirens populations with CMB prior information in order to have a forecast of how much this combination can improve the precision on the parametric space. Taking LISA + CMB prior, we note a 20\%, 33\% and 15\% accuracy from the Pop III, Delay and No Delay populations, respectively. Moreover, $H_0$ can be simultaneously fit to 0.8\%, 1.2\% and 0.7\% accuracy from the Pop III, Delay and No Delay populations, respectively. Note that the cases Delay and No Delay populations, generate higher and lower values for $\Omega_{m}$, respectively, when compared  with the value observed by CMB. On the other hand, the parameter $\xi$ can not be measured via electromagnetic waves probes as discussed earlier. Any possible improvements on $\xi$  will happen by breaking the degeneracy on the other parameters, in special $\Omega_{m}$, because $\xi$ and $\Omega_{m}$ are correlated with each other. When we add the CMB prior, the pair ($\Omega_{m}$, $H_0$) tends to (0.31, 67.4). Looking at the parametric space $\xi-\Omega_{m}$ (see Fig. \ref{LISA_II}, \ref{LISA_III}), we can notice this effect. Therefore, the compression on $\Omega_{m}$ to Planck-values, can increase the relative error on $\xi$ in some cases. We note it more significantly in the Delay case. A way to robustly improve the total constraint on $\xi$ would be to combine our analysis with weak lensing and redshift-space distortions measurements. Because in this case, all probes are very sensitive to $\xi$ effects. These steps will be left for future perspectives.
\\

We have extended the analysis in Ref.~\citet{Mitra:2020vzq} 
considering mock data from the LISA perspective.
We see an overall improvement on the precision of the Hubble constant, $H_0$, and $\Omega_m$ by including the CMB information. However, we see that by including the CMB it is  difficult to distinguish the model from GR as the 68\% credible intervals on $n$ and $\xi$ in  tables \ref{tab:main_results_1}, \ref{tab:main_results_2} and \ref{tab:main_results_3} contain the GR prediction, namely $\xi=1$. We note the prospective strength of implementing the CMB is that the degeneracy between  $\xi$ and $\Omega_m$  can be broken.

\section{Final remarks}\label{5}

We have presented a forecast analysis for the standard siren events which are expected to be observed in LISA frequency band for  three categories of population models, named, Pop III, Delay and No Delay. Then, we performed a parameter estimation analysis to test modifications of GR inspired by the No Slip gravity, where the speed of GWs propagation is equal to the speed of light, and  the effective gravitational coupling strengths to matter and light are equal, but yet different from Newton’s constant coupling.

In our analysis we have also included the priors from the CMB and find that perspectives towards No Slip gravity can be well tested within the expected perspective for cosmological observations with LISA. Adding the CMB information could help improve our parameter estimation and the shape of covariances in the parameter space. In combination with CMB information, we find a 15\% accuracy on the modified gravity free parameters and 0.7\% accuracy on the Hubble parameter.

In this present work we have considered one space-borne detector. A more sophisticated future study will be to include both LISA and TianQin or Taiji \citet{Ruan:2018tsw}. This could increase  
the number of the GW events with electromagnetic counterparts and improve the whole parameter space of the model. In the next decades GW detectors will form a powerful detector network. This will allow to do joint detections for  events. This can be utilized to do complementary studies of modified gravity theories with modified GW propagation equations. On the other hand, the modified GW propagation arises naturally from any modified gravity theory. It may be interesting to search and improve the expected study of LISA sources without electromagnetic counterpart, and use these information to test modified GW propagation, since several LISA sources are expected at large cosmological distance \citet{Mukherjee:2020mha,Borhanian:2020vyr,Wang:2020dkc,Feeney:2020kxk}. Thus, we can significantly increase our sample, number of events for doing statistics. Also, it may be interesting to search the modified GW propagation effects on the gravitational wave background signal expected in LISA band \citet{DOrazio:2018jnv,Samsing:2018isx, Kowalska}.

\section*{Acknowledgments}

The authors thank the referee for some useful comments that improved the manuscript. R.C.N. acknowledges financial support from the Funda\c{c}\~{a}o de Amparo \`{a} Pesquisa do Estado de S\~{a}o Paulo (FAPESP, S\~{a}o Paulo Research Foundation) under the project No. 2018/18036-5. DFM thanks the Research Council of Norway for their support. Some computations were performed on resources provided by UNINETT Sigma2 -- the National Infrastructure for High Performance Computing and Data Storage in Norway.
\section*{Data Availability}
The data underlying this article will be shared on reasonable request to the corresponding author.
\bibliographystyle{mnras}
\bibliography{lisa}

\bsp	
\label{lastpage}
\end{document}